\documentclass[preprint2]{aastex}

\newcommand{\ebv}{\mbox{$E(\bv)$}}
\newcommand{\vi}{\mbox{$V\!-\!I$}}
\newcommand{\jh}{\mbox{$J\!-\!H$}}
\newcommand{\nicmos}{{\small NICMOS}}
\newcommand{\dbump}[1]{\mbox{${\Delta #1}_{\rm HB}^{\rm bump}$}}
\newcommand{\dmhbto}[1]{\mbox{${\Delta #1}_{\rm HB}^{\rm TO}$}}
\newcommand{\daoall}{{\small DAOPHOT/ALLSTAR}}
\newcommand{\allstar}{{\small ALLSTAR}}
\newcommand{\msun}{\mbox{${\cal M}_{\sun}$}}
\newcommand{\pc}{\mbox{pc}}
\newcommand{\kpc}{\mbox{kpc}}
\newcommand{\zzin}{\mbox{${\rm [Fe/H]_{Z85}}$}}

\newcommand{\beq}{\begin{equation}}
\newcommand{\eeq}{\end{equation}}

\makeatletter
\newenvironment{tablehere}
  {\def\@captype{table}}
  {}

\newenvironment{figurehere}
  {\def\@captype{figure}}
  {}
\makeatother

\shorttitle{HST-NICMOS observations of Terzan~5}
\shortauthors{Cohn et al.}

\begin{document}
\title{HST-NICMOS Observations of Terzan~5: \\ 
	Stellar Content and Structure of the Core}

\author{Haldan N. Cohn and Phyllis M. Lugger}
\affil{Department of Astronomy, Indiana University, 
       727 E. Third Street, Bloomington, IN  47405}
\email{cohn@indiana.edu}

\and

\author{Jonathan E. Grindlay and Peter D. Edmonds}
\affil{Harvard College Observatory, 60 Garden Street, Cambridge, MA 02138}

\begin{abstract}

We report results from HST-\nicmos\ imaging of the extremely dense
core of the globular cluster Terzan~5.  This highly obscured bulge
cluster contains a low-mass X-ray binary and at least two millisecond
pulsars, with evidence for a large millisecond pulsar population.  It
has been estimated to have one of the highest collision rates of any
galactic globular cluster, making its core a particularly conducive
environment for the production of interacting binary systems.

We have reconstructed high-resolution images of the central
$19\arcsec\times19\arcsec$ region of Terzan~5 by application of the
drizzle algorithm to dithered NIC2 images in the F110W, F187W, and
F187N near-infrared filters.  We have used a \daoall\ analysis of
these images to produce the deepest color-magnitude diagram (CMD) yet
obtained for the core of Terzan~5.  We have also analyzed the parallel
$11\arcsec\times11\arcsec$ NIC1 field, centered $30\arcsec$ from the
cluster center and imaged in F110W and F160W, and an additional NIC2
field that is immediately adjacent to the central field.

This imaging results in a clean detection of the red-giant branch and
horizontal branch in the central NIC2 field, and the detection of
these plus the main-sequence turnoff and the upper main sequence in
the NIC1 field.  We have constructed an $H$ versus \jh\ CMD for the
NIC1 field, which provides a measurement of the infrared reddening,
$E(\jh) = 0.72$ (corresponding to $E(\bv) = 2.16$).  We obtain a new
distance estimate of 8.7~kpc, which places Terzan~5 within less than
1~kpc of the galactic center.  The NIC1 CMD provides an apparent
detection of the red giant branch bump, about 1 mag below the
horizontal branch, indicating that the metallicity is at least solar
and possibly somewhat higher.  A number of blue objects are detected
in the central NIC2 field and the NIC1 field, which hint at the
presence of a blue horizontal branch.  We have also determined a
central surface-density profile which results in a maximum likelihood
estimate of $7\farcs9\pm0\farcs6$ for the cluster core radius.

We discuss the implications of these results for the dynamical state of
Terzan~5. 

\end{abstract}

\keywords{globular clusters: individual 
(Terzan 5)---Hertzsprung-Russell diagram---infrared: stars \\ 
\vbox{\vspace*{0.5in}
To appear in the {\sl The Astrophysical Journal}, May 20, 2002}
}

\section{Introduction \label{intro}}

The extremely dense core of the bulge globular cluster Terzan~5 makes
it a likely location for the dynamical production of interacting
binaries.  It has been estimated to have the highest stellar
interaction rate of any galactic globular cluster \citep{ver87}.
Indeed, Terzan~5 is one of only 12 clusters known to contain a
low-mass X-ray binary \citep*[LMXB;][]{joh95}; this object exhibits
both transient behavior and bursts.  Terzan~5 also contains an
eclipsing millisecond pulsar \citep[MSP;][]{nic92}, many
steep-spectrum radio point sources within 30\arcsec\ of the cluster
center, and steep-spectrum extended radio emission in the cluster core
\citep{fru95}.  \citet{lyn00} have recently detected a second MSP just
10\arcsec\ from the cluster center.  Based on the unresolved radio
flux from the core of Terzan~5, \citet{fru00} argue that it may have
the largest pulsar population of any cluster in the Galaxy.  They also
detect three discrete radio sources that they interpret as likely
pulsars.

It has long been thought that dynamical interactions among stars in
dense environments may lead to the production of interacting binaries
and/or merged stars.  Interaction mechanisms include two-body tidal
capture and various three-body interactions, such as exchange reactions
in which a massive degenerate star may replace one member of a
main-sequence binary.  This picture is supported by Hubble Space
Telescope (HST) studies, which have made the first detections of
cataclysmic variables (CVs) in cluster cores \citep[e.g.\ in
NGC~6397;][]{coo95} and have also detected substantial, centrally
concentrated blue straggler populations \citep[e.g.\ in M80;][]{fer99a}. 

Ground-based study of Terzan~5 has been hampered by its high
obscuration \citep*[$A_V \approx 7.7$;][]{ort96} and the extreme
crowding in its core.  Given its galactic coordinates, ($\ell =
3\fdg8, b=1\fdg7$), the line of sight to Terzan~5 passes within
0.6~kpc of the galactic center.  \citet{ort96} produced the first
color-magnitude diagram (CMD) for Terzan~5, from imaging obtained
under conditions of exceptional seeing (0\farcs3 -- 0\farcs5) at the
New Technology Telescope (NTT).  Nevertheless, their $I$ vs.\ $(V-I)$
CMD reaches only to just below the horizontal branch (HB) in the
central 0\farcm5 radius about the cluster center.  Moreover, their CMD
is strongly affected by differential reddening, with the red giant
branch (RGB) spanning nearly a magnitude in color.  Based on their
CMD, \citet{ort96} estimate that the metallicity of Terzan~5 is
approximately solar, making it one of the most metal rich of all
globular clusters.  This is consistent with the range of earlier
estimates of ${\rm [Fe/H]} = +0.24$, based on integrated infrared
photometry \citep{zin85}, ${\rm [Fe/H]} = -0.28$, based on integrated
spectroscopy \citep{arm88}, and ${\rm [Fe/H]} = +0.07$, also based on
integrated spectroscopy \citep{bic98}.  \citet{ort96} also obtained a
distance estimate of $d = 5.6~{\rm kpc}$ from their CMD, suggesting
that Terzan~5 may be significantly closer than previous estimates,
which generally lie in the range $d = 7-9~\kpc$.  Given its galactic
coordinates, this range allows the possibility that Terzan~5 is
located very close to the galactic center.

Terzan~5 is a natural target for HST-\nicmos\footnote{\nicmos\ = Near
Infrared Camera and Multi Object Spectrometer} imaging studies, both
to better characterize the stellar content and structure of the
central regions, and to search for additional evidence of the presence
of interacting binaries there.  We report here results from \nicmos\
observations of the central region of Terzan~5.  We describe the
observations in \S2, present the analysis in \S3, and discuss our
conclusions in \S4.  This paper concentrates on reporting the analysis
of the cluster CMD and central structure.  A companion paper reports
the results of a search for variable stars and the counterparts of the
bright LMXB and the eclipsing MSP \citep{edm01}.  While the present
paper was in the final stages of the refereeing process, an
independent \nicmos\ study of the CMDs of several reddened bulge
clusters, including Terzan~5, appeared in print \citep{ort01}.  Their
study focuses on cluster age determination by isophote fitting; we
compare our results for the age of Terzan~5 with theirs in
\S\ref{age}.

\section{Observations \label{observations}}

The primary targets for our \nicmos\ observations were the central
region of Terzan~5 and the MSP position 0\farcm5 from the center
\citep{edm01}.  We imaged Terzan~5 with two primary pointings, one in
which NIC2 was centered on the cluster center and one in which NIC1
was centered on the MSP position.  These observations produced
parallel NIC1 observations at 30\arcsec\ from the cluster center and
parallel NIC2 observations of a field adjacent to and slightly
overlapping with the central field.  Figure~\ref{fig:map} shows the
three fields used in the present study, viz.\ the central and adjacent
NIC2 fields, and the NIC1 field that was obtained in parallel with the
central field.  The exposures used in this study include: (1) eight
exposures of the central NIC2 field in each of the filters F110W,
F187W, and F187N, (2) eight exposures of the NIC1 field in each of the
filters F110W, F160W, and F187N, and (3) two exposures of the offset
NIC2 field in F187W\@.  For the exposures of the NIC1 and central NIC2
fields, the durations were 320\,s for the broad-band (W) filters and
1536\,s for the narrow-band (N) filter.  For the offset NIC2 field,
the total exposure time was 832\,s.

\begin{figurehere}
\centering\includegraphics[width=3.2in]{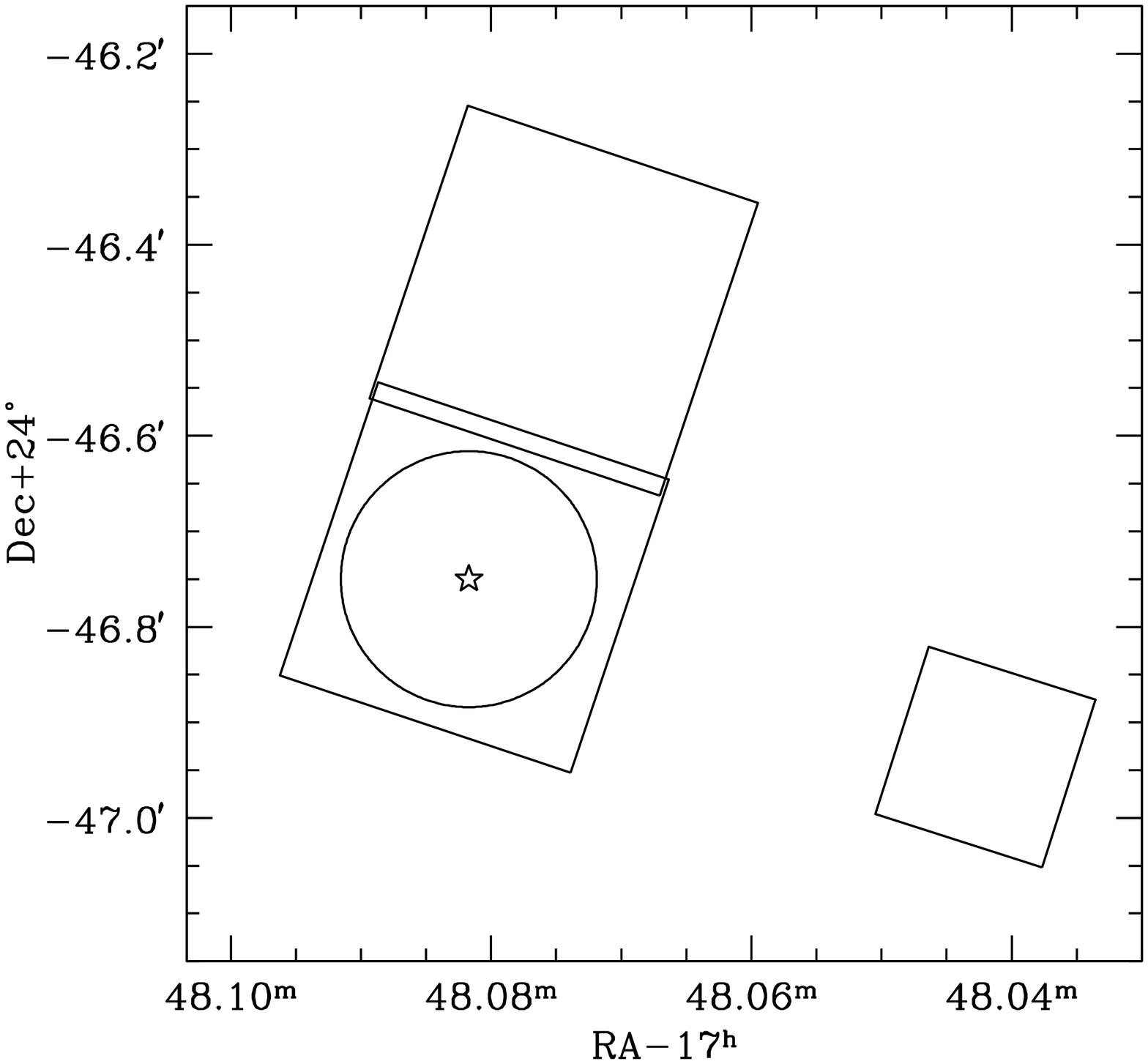}
\vspace*{-0.35in}
\figcaption[f1.eps]{\protect\small Map of the three
\nicmos\ fields used in this study.  North is up and east is to the
left.  The star indicates our determined center of Terzan~5 and the
circle indicates our maximum-likelihood determination of the core
radius.  The two eastern fields were observed with the NIC2 (field
width = 19\farcs2), while the smaller western field was observed with
the NIC1 (field width = 11\arcsec).
\label{fig:map}}
\bigskip
\end{figurehere}

The eight observations of the central NIC2 field and the parallel NIC1
field were dithered, in order to increase the effective angular
resolution, and consequently the photometric accuracy, in this
extremely crowded field.  The F187W and F187N filters, which are
centered on Paschen-$\alpha$, were chosen for identifying
emission-line objects.  The F110W filter (which approximately
corresponds to the Johnson $J$ band) was included to produce a
broad-band CMD, in combination with the F187W filter.  Since F187W is
not available for NIC1, we instead used F160W (which approximately
corresponds to the Johnson $H$ band) for the redder broad-band filter
for this camera.  While this has the disadvantage of providing a
somewhat different CMD for the NIC1 field, a substantial advantage is
that it allowed us to construct a NIC1-field CMD in a standard
ground-based filter system ($J$ and $H$).  This proved to be quite
valuable in allowing us to determine new values for the metallicity,
reddening, and distance of Terzan~5, as described below.

The eight dither positions for the NIC2 field were selected in a
$3\times3$ array (minus one corner position).  The basic step between
dither positions was 3 NIC2 pixels (0\farcs23), with a random 0.125
NIC2 pixel (0\farcs009) step added to provide subpixel sampling.  Each
dither position was observed for one HST orbit.  The 8-orbit
observation sequence was split over two visits, separated in time by
one month.  Cross correlations of the resulting images indicate that
subpixel pointing was accurately maintained from orbit to orbit within
each visit.  However, there were pointing shifts by about 0.5 and 0.2
NIC2 pixels in the two orthogonal directions between the visits.
These additional shifts acted in concert with the intentional 0.125
pixel shifts to provide subpixel sampling.

\section{Data Analysis}

\subsection{Drizzle Reconstruction of Images}

The extraordinarily crowded central NIC2 field presents a substantial
challenge to photometric analysis.  The NIC1 and NIC2 provide full
Nyquist sampling (i.e.\ 2 pixels per resolution element) down to
1\micron\ and 1.7\micron, respectively.  Thus the NIC2 images in the
F110W band are strongly undersampled.  The effect of this
undersampling is alleviated by our dithering strategy for the central
region.  In order to make optimal use of the information contained in
the dithered images, we employed the ``drizzle'' (variable-pixel
linear reconstruction) algorithm of \citet{hoo97} to recover the full
resolution provided by the HST optics.  We used the implementation of
this algorithm from the STSDAS extensions of IRAF\footnote{IRAF is
distributed by the National Optical Astronomy Observatories, which are
operated by the Association of Universities for Research in Astronomy,
Inc., under cooperative agreement with the National Science
Foundation.}, as described by \citet{hoo99}.

For the reconstructed NIC1 and NIC2 images, we chose pixel sizes of
0\farcs022 and 0\farcs038, respectively, which are half of the
original pixel sizes.  We chose a value of 0.5 input pix for the
``drop'' size; this parameter determines the magnification of the
projection of each input pixel into the output image.  Examination of
the output drizzled images indicated a moderate level of residual
pixel-to-pixel fluctuation, suggesting incomplete PSF sampling.  These
fluctuations were particularly evident in the vicinity of stellar
image cores in the residual image on performing PSF-subtraction
photometry.  We reduced this noise by applying a Gaussian smoothing
filter ($\sigma = 0.75~\mbox{pix}$) to the drizzled images.
Figure~\ref{fig:dither_comp} shows a comparison of a single F110W
exposure of the central region and the result of combining the eight
F110W images using the drizzle algorithm.  

\begin{figurehere}
\bigskip
\centering\includegraphics[width=3.0in]{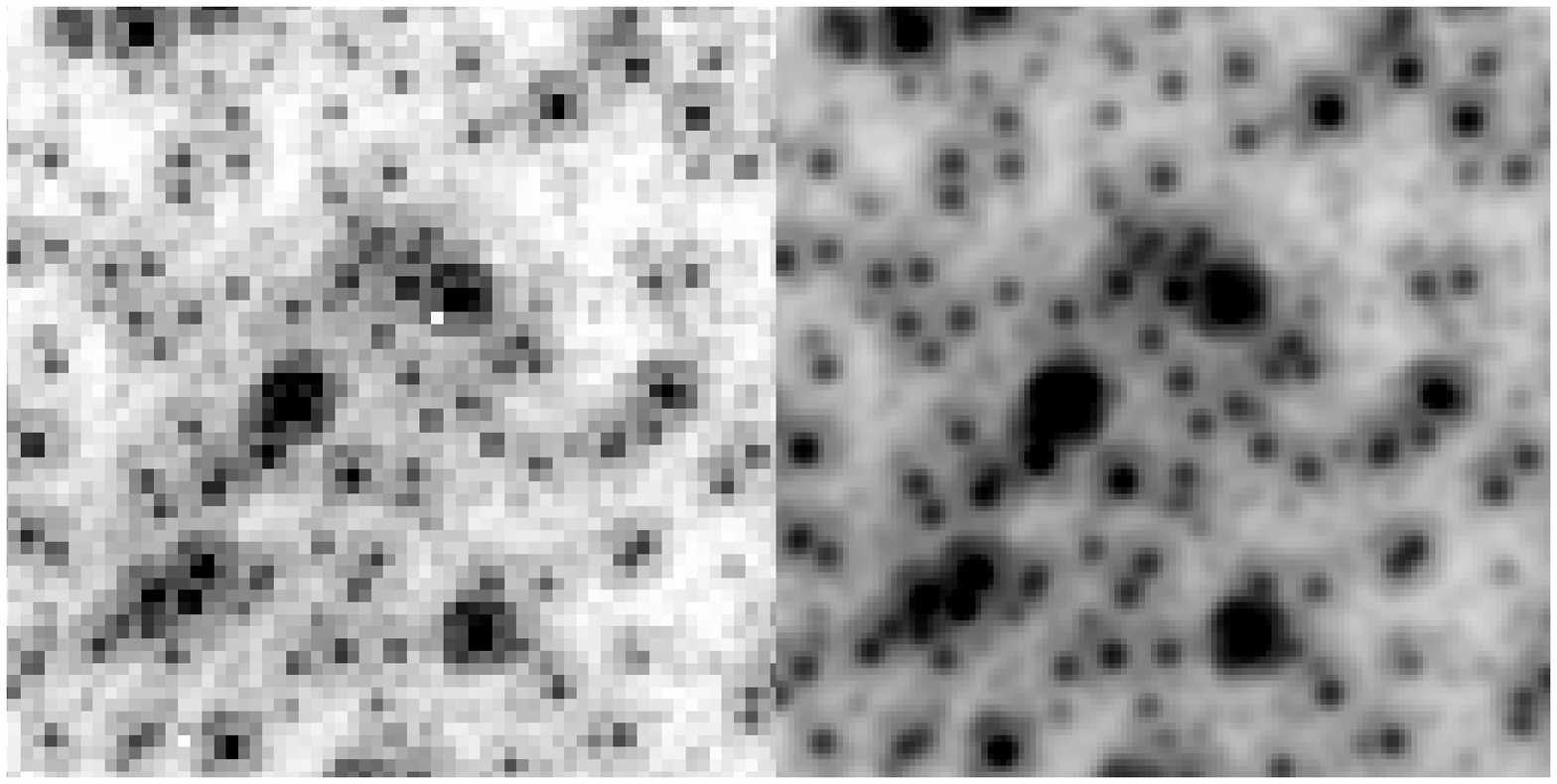}
\figcaption[f2.eps]{\protect\small Comparison of the central
$5\arcsec\times5\arcsec$ section of a single F110W NIC2 exposure of
the cluster center (left panel) with the corresponding section of the
image that results from a drizzle reconstruction from eight dithered
images (right panel).
\label{fig:dither_comp}}
\bigskip
\end{figurehere}

\subsection{Photometry}

Given the substantial increase in the resolution of the PSF that
results from the application of the drizzle algorithm, it is feasible
to carry out PSF-subtraction photometry.  We employed the \daoall\
software package \citep{ste87,ste88} to determine an empirical PSF for
each frame and simultaneously fit this to all detected stars.
Approximately 100--150 stars were used to determine a PSF with
quadratic spatial variation across the frame.  A complication in the
process lies in distinguishing between faint stars and artifacts of
the complex PSF structure of bright stars.  We used visual inspection
of apparent detections in the vicinity of bright stars to help filter
out artifacts.  Since the diffraction pattern has a strong wavelength
dependence, inter-comparison of the F110W and F160W images proved
useful in this filtering process.  A PSF radius of 20 pixels was used
for both the NIC1 and NIC2 fields.  This size was chosen as a
compromise between the goals of subtracting as much of the halos of
bright stars as possible, while still retaining a sufficient number of
stars, with no brighter neighbors within the PSF radius, to accurately
characterize the spatial variation of the PSF\@.

Given the higher degree of crowding in the central NIC2 field,
relative to the offset NIC1 field, and the nearly twice as large NIC2
pixels, PSF subtraction was less effective in the NIC2 field.  A clear
indication of this is that the limiting magnitude is considerably
fainter in the NIC1 field for the same amount of exposure (see
\S\ref{cmd}).  Also, examination of the residual frames for the NIC1
and NIC2 fields indicates a considerably greater amount of residual
light in the latter.  Nevertheless, PSF-subtraction photometry
produced significantly better results than simple aperture photometry,
for both the NIC1 and NIC2 frames.

\subsection{Calibration}

The \nicmos\ pipeline calibration system produces images that are
calibrated in flux units, i.e.\ the images are given in count rate
(DN~s$^{-1}$).  These can be converted to physical flux units (Janskys
or erg~s$^{-1}$~cm$^{-2}$~\AA$^{-1}$) by multiplying by a corresponding
scale factor that is tabulated for each filter.  These fluxes can then
be expressed in magnitude units in the AB$_\nu$ or Vega systems; we use
the latter.  As noted previously, the F110W and F160W filters, which
were used for the NIC1 field, are analogs to the ground-based $J$ and
$H$ bands.  We computed a linear transformation between the HST and
ground-based systems using \nicmos\ photometry for five standard stars
with a wide color range ($-0.1 \le \jh \le +2.1$) that are listed in the
\nicmos\ Data Handbook, Version 4.0.  Preliminary \nicmos\ photometry
for the stars is given at the \nicmos\ website.  The resulting
transformations are:
\beq
J = F110W + 0.033 - 0.335\, (F110W-F160W)  \label{eqn:J_trans}
\eeq
\beq
H = F160W - 0.001 - 0.092\, (F110W-F160W)  \label{eqn:H_trans}
\eeq
These results are almost identical to those of \citet{sch99} who
followed a similar calibration procedure.  The RMS residual is
0.16~mag for the $J$-transformation and 0.05~mag for the
$H$-transformation.  The large value of the former is mostly
determined by one large residual (for the red standard BRI0021 with
$J-H=1.17$); the RMS of the other four $J$-residuals is 0.07~mag.  We
also note the existence of a significant color term in the
$J$-transformation, which also contributes to the overall uncertainty.

The transformations given by equations~(\ref{eqn:J_trans}) and
(\ref{eqn:H_trans}) allowed us to construct a standard near-infrared CMD
for the NIC1 field.  Since F187W (rather than F160W) serves as the
redder broad-band filter in the central NIC2 field, it is necessary to
use a color index for this field that does not have a strict
ground-based analog.

\section{The Color-Magnitude Diagram \label{cmd}}

In constructing CMDs, we required that the centroids of stars
determined by \allstar\ match to 0.5 pix between the F110W and F187W
frames.  This tight limit on the positional offset helped to filter
out both weakly detected stars with uncertain photometry and PSF
artifacts.  We also eliminated stars that have neighbors with
comparable magnitudes and centroids within 3 pixels, since these were
strongly blended and thus produced substantial photometric uncertainty.

Figure~\ref{fig:cmd_comp} shows a comparison of the CMDs obtained for
the NIC2 and NIC1 fields.  The ordinate is F110W in both cases.  It is
immediately obvious that the central NIC2 field is much more populous,
owing to the much higher stellar density and the four times larger
area.  Not surprisingly, the offset NIC1 field produces much deeper
photometry, given the lower crowding and the smaller pixel size.  The
horizontal branch (HB) is clearly detected in both fields at a
magnitude of $\mbox{F110W} \approx 15.9$, demonstrating the
consistency of the calibration between the NIC1 and NIC2 cameras.  The
NIC1 photometry also clearly detects the main sequence turnoff (MSTO)
at $\mbox{F110W} \approx 20$ and upper main sequence; the faint-end
limit of the NIC2 photometry is brighter than this at $\mbox{F110W}
\approx 19$.

\begin{figurehere}
\bigskip
\centering\includegraphics[width=3.2in]{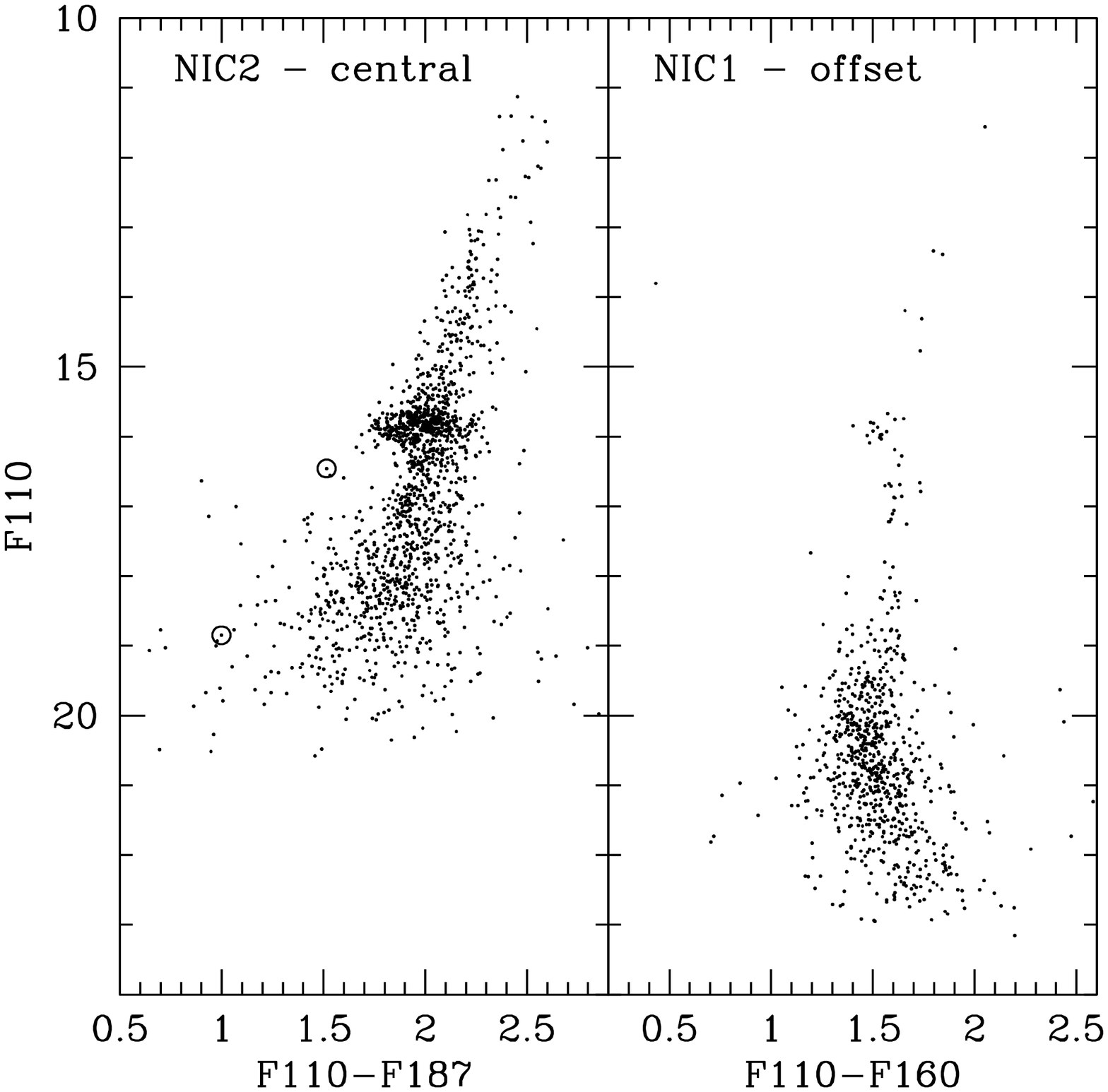}
\figcaption[f3.eps]{\protect\small Comparison of the CMDs for the
central NIC2 field (left panel) and the NIC1 field (right panel).  The
ordinate is F110W in both cases, while the abscissa is the available
broad-band color index.  The much greater depth of the NIC1 CMD is a
consequence of the much lower crowding and $2\times$ higher resolution
in this field.  The circled objects in the NIC2 field are
photometrically variable \citep{edm01}; the brighter object is an RRab
Lyrae and the fainter one is likely an eclipsing blue straggler.
\label{fig:cmd_comp}}
\bigskip
\end{figurehere}

While one of the original goals of this study was to reach below the
MSTO in the central NIC2 field, in order to survey the core of Terzan~5 for
CVs, this is clearly not possible.  Even with extensive dithering, the
extraordinary crowding of this region substantially limits the accuracy
of the photometry, particularly for $\mbox{F110W}>18$, which is just two
magnitudes below the HB\@.  Nevertheless, the photometry in the central
NIC2 field substantially supersedes the quality of the excellent
ground-based $V$ and $I$ photometry reported by \citet{ort96}.
Examination of their CMD for the central $26\arcsec$ radius region
indicates that the limiting magnitude is not much more than one
magnitude below the HB and that the width of the RGB is
$\Delta(\vi)=0.92$; they ascribe this to differential reddening.  The
greatly increased depth and precision of the CMD that is afforded by
our NICMOS observations permits us to search for unusual blue
objects below the HB, if not below the MSTO\@.  Furthermore, as has been
recently demonstrated by \citet{dav00,dav01}, near-infrared CMDs provide
valuable information on the metallicities, reddenings, and distances of
low-latitude globular clusters. 

In addition to constructing broad-band CMDs, we also constructed a
$\mbox{F187N}-\mbox{F187W}$ versus F187W diagram to search for evidence
of emission-line objects.  We find no convincing detections of such
objects in either the central NIC2 field or the NIC1 field, in agreement
with the analysis of the same data set by \citet[][see their
Fig.~8b]{edm01}.

\subsection{Blue Stars below the HB \label{blue_stars}}

Figure~\ref{fig:cmd_comp} indicates a number of stars that lie
blueward of the RGB in the NIC2 field.  \citet{edm01} have shown that
two of these blue stars show significant variability; these are
circled in the left panel of Figure~\ref{fig:cmd_comp}.  As discussed
by \citet{edm01}, the brighter of these variables is an RRab Lyrae,
while the fainter is likely an eclipsing blue straggler or possibly
the LMXB counterpart.  These two stars are both located at $8''$ from
the cluster center.  This puts the objects at the projected core
radius.  While the possibility of a chance superposition always
exists, particularly in a crowded field, the small radial offset of
these stars from the cluster center makes a superposition of a
nonmember unlikely.  Moreover, given the consistency of the apparent
magnitude of the RR~Lyrae with the value expected for one located in
Terzan~5, it appears likely that the star is at the cluster distance.
While the photometry of the RR~Lyrae appears fairly secure, the
candidate blue straggler is clearly in the region where photometric
uncertainty substantially widens the RGB\@.  No variability is
observed from the other two stars that lie close to the RR Lyrae in
the CMD\@.

Given the extreme crowding of the central field of Terzan~5, imperfect
PSF-subtraction photometry can produce large photometric error as well
as outright artifacts, e.g.\ PSF ``tendrils'' that are detected as
stars.  We performed several tests of the photometry of the apparently
blue stars, including visual inspection of the detections in the
drizzled frames, variations in the photometric procedure, and cuts on
the magnitude error estimates provided by \allstar.  Visual inspections
generally proved useful at and above the HB level, where photometric
problems were fairly clear.  However, by about 1 mag below the HB, it
became quite difficult to discern photometric problems in this way, due
to the extreme crowding of the field and the complexity of the PSF\@.
Thus, we deemed automated photometric tests to be more useful for faint
stars.  To this end, we investigated cuts on the formal magnitude error
estimates provided by \allstar\@.  The result of one such experiment is
shown in Figure~\ref{fig:nic2cmd}.  In this case, a limit of 0.03 was
placed on the formal errors of the F110W and F187W magnitudes.  The
three bluest stars are most likely foreground objects; a strong sequence
of such objects is seen in the $2\farcm2\times2\farcm2$ field surveyed
by \citet{ort96}. 

\begin{figurehere}
\bigskip
\centering\includegraphics[width=3.2in]{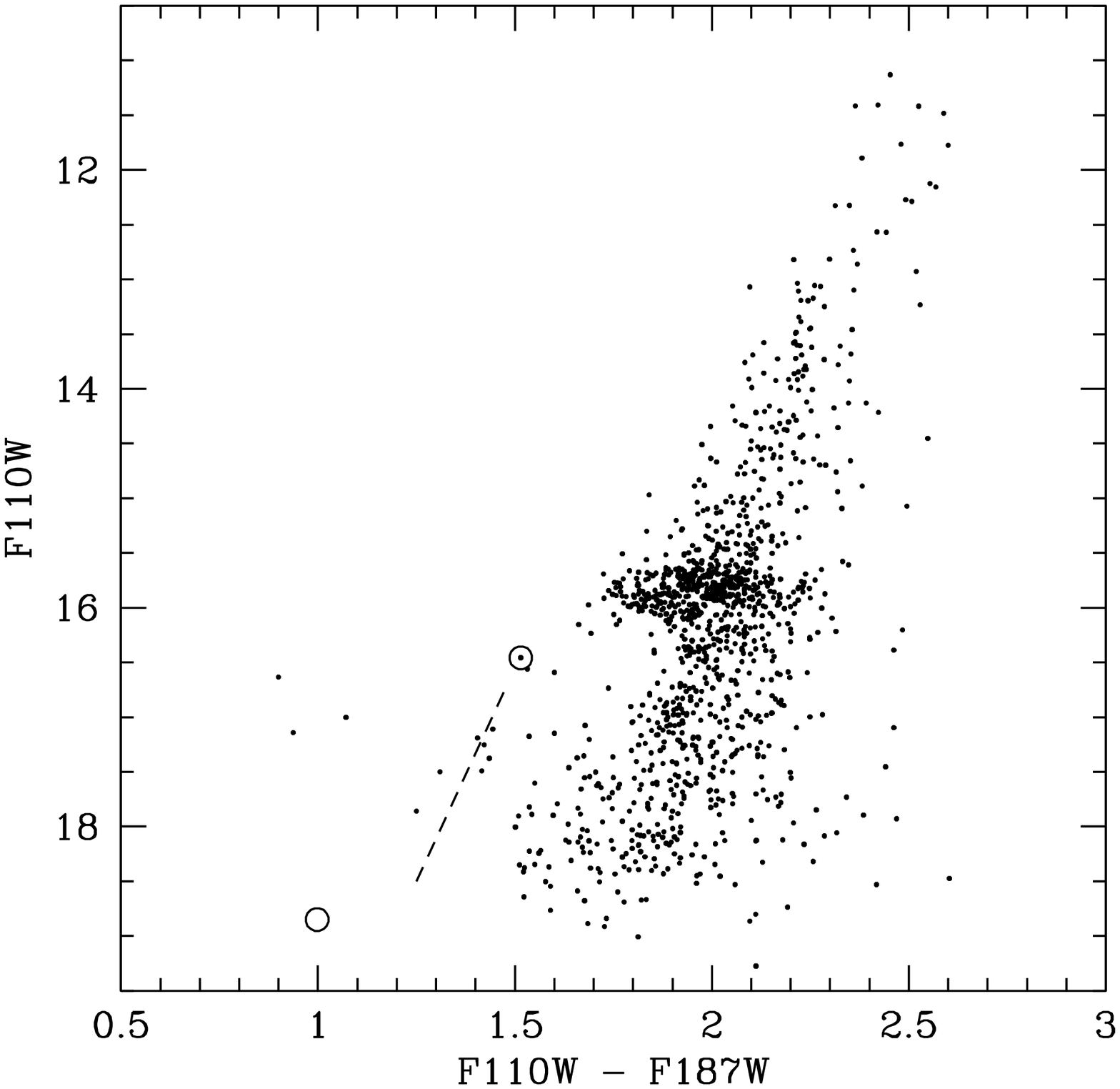}
\figcaption[f4.eps]{\protect\small CMD for the
central NIC2 field.  The photometry is the same as in the left panel
of Fig.~\ref{fig:cmd_comp}, with a 0.03 mag limit placed on the formal
photometric error.  The circled objects are the variables noted in
Fig.~\ref{fig:cmd_comp}; the fainter one does not pass the magnitude
error filter.  The dashed line indicates the location of the possible
blue horizontal branch.  The three bluest stars, near F110W = 17 are
probably foreground stars.
\label{fig:nic2cmd}}
\end{figurehere}

\newpage

Figure~\ref{fig:nic2cmd} provides a hint of the presence of a blue
horizontal branch (BHB) that is parallel to the RGB and about 0.5 mag
bluer in color.  This is the expected location of the BHB in infrared
colors, as can be seen in the photometry of four metal-poor clusters
by \citet{dav99}.  The hint of a BHB in Terzan~5 is intriguing, since
the presence of a BHB in an extremely metal rich cluster is unusual.
\citet{ric97} first discovered substantial BHB populations in old,
metal-rich systems.  They found a blue extension of the strong red
horizontal branch (RHB) in the globular clusters NGC~6388 and
NGC~6441, based on HST WFPC2 $BV$ imaging.  These clusters have high
[Fe/H] values of $-0.60$ and $-0.53$, respectively.  Both clusters
also have extremely high central densities, with $\log \rho_0
~(\msun\,\pc^{-3})$ values of 5.3 and 5.4, respectively \citep{djo93}.
\citet{ric97} suggest that the presence of a BHB in these clusters may
be the result of enhanced mass loss from giants induced by stellar
interactions.  However, they note that the BHB/RHB ratio does not show
the radial gradient that might be expected if BHB stars were produced
by dynamical interactions in the cluster cores.  Moreover,
\citet{bed00} have shown that the extended BHB population in the
cluster NGC~2808 is no more centrally concentrated than the RGB
population.  A complicating factor is that once a star loses a
substantial amount of envelope mass, by either stellar interactions or
normal giant winds, it will tend to move to larger orbital radius as a
consequence of mass segregation.  Given HB lifetimes of $\sim
10^8$~yr, this mass segregation process has enough time to at least
partially reverse an initial central concentration of BHB stars.  

Since Terzan~5 has an even higher central density \citep[$\log \rho_0
= 5.8$;][]{djo93} than do NGC~6388 and NGC 6441, dynamical interaction
effects should be even more important for Terzan~5.  Given the handful
of potential BHB stars detected by our photometry, it is not feasible
to examine the radial dependence of the BHB/RHB ratio.  This will
require wide-field infrared imaging of Terzan~5.

\subsection{The RGB Bump and Metallicity}

Stellar evolution theory predicts the presence of a RGB ``bump,'' a
pileup that results from a pause in the ascent of the RGB that occurs
when the hydrogen burning shell extends past the point of last maximum
depth of the convective envelope \citep[see e.g.][]{ibe68}.
\citet{fus90} demonstrated that the location of the bump on the RGB is
correlated with the cluster metallicity.  \citet{fer99b} have measured
the location of the RGB bump in 47 globular clusters and have
calibrated the relation between the offset of the bump magnitude from
the HB magnitude ($\dbump{V}\equiv V_{\rm bump} - V_{\rm HB}$) and
several different metallicity indices.  The bump magnitude is brighter
than the HB for metallicities below ${\rm [Fe/H] \approx -1.2}$ and is
fainter than the HB for higher metallicities.  The clusters in the
\citet{fer99b} survey range in [Fe/H] from $-2.2$ to $-0.2$. The high
end of this range is of particular interest, for comparison with
Terzan~5.  The most metal rich cluster in the sample is NGC~6528 with
${\rm [Fe/H]} = -0.23$ on the \citet{zin85} scale and $\dbump{V} =
+0.78$.  \citet{dav00} has recently reported near infrared photometry
of NGC~6528, with the Canada-France-Hawaii Adaptive Optics system, and
finds a similar bump offset in the infrared, viz.\ $\dbump{K}=+0.6$.

\citet{fer99b} and \citet{dav00} note that examination of the cluster
luminosity function provides a good means of determining the magnitude
of the RGB bump.  The top and middle panels of
Figure~\ref{fig:lf_comp} show F110W luminosity functions for NIC2 and
NIC1 fields, respectively.  The results for the NIC2 frame are shown
for both the entire frame and the corners of the frame only.  The
motivation for separately considering the corners is to reduce the
influence of photometric error, which is greatest in the central part
of the frame owing to the greater crowding there.  The bottom panel
shows the ratio of the NIC2 and NIC1 luminosity functions, which
provides a measure of the completeness of the counts in the NIC2
field.  Only bins containing at least three stars are shown.  The
solid horizontal line shows the overall count ratio for $\mbox{F110W}
< 18.4$.  The NIC2/NIC1 count ratio shows that significant
incompleteness sets in in the NIC2 field for $\mbox{F110W} > 18.0$;
this is about 2 mag fainter than the HB\@.  This incompleteness is not
surprising, given the extraordinarily high central density of
Terzan~5, together with the complex \nicmos\ PSF structure.

\begin{figurehere}
\bigskip
\centering\includegraphics[width=3.2in]{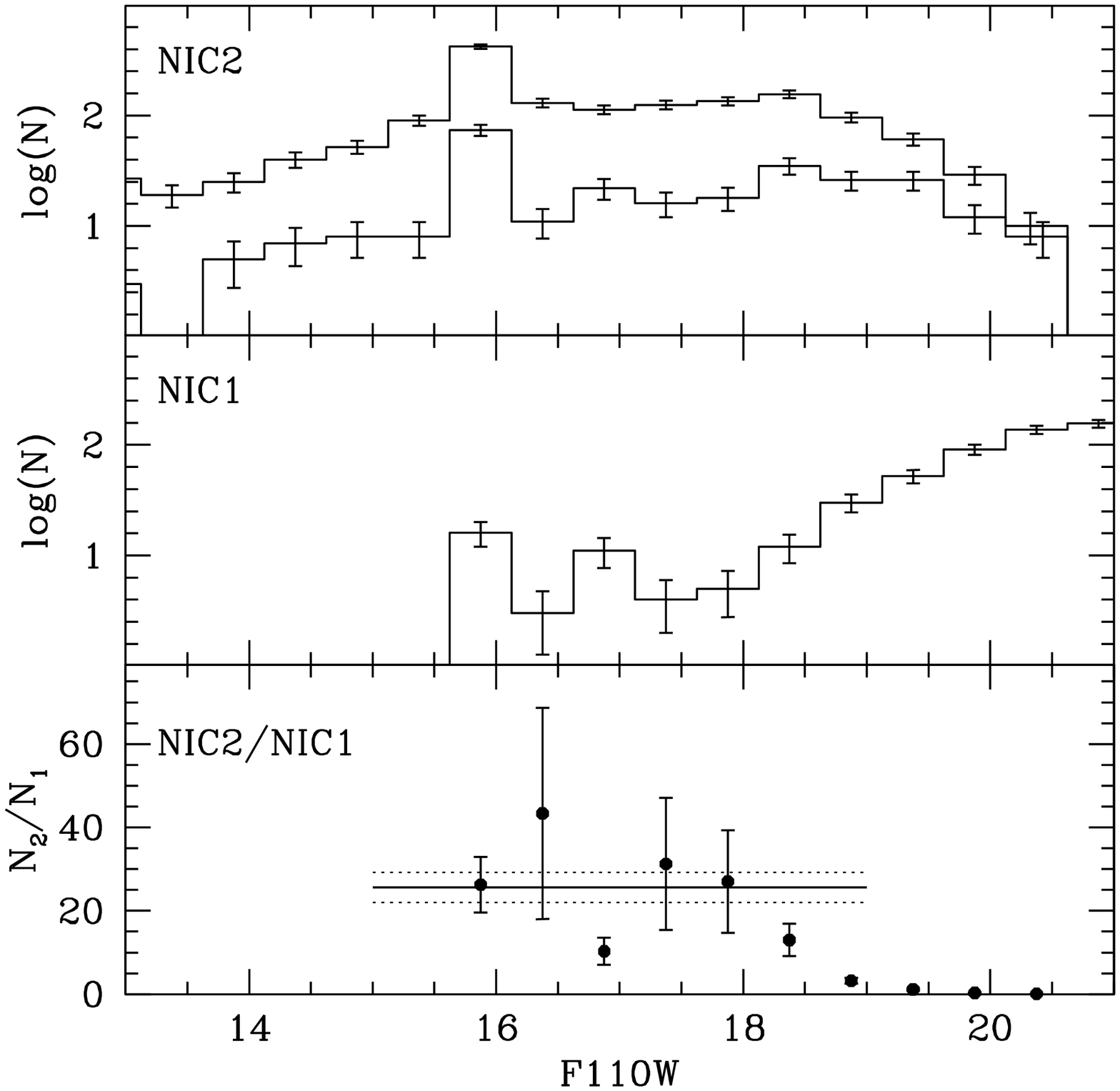}
\figcaption[f5.eps]{\protect\small Comparison of the
luminosity functions for the central NIC2 field (top panel) and the
NIC1 field (middle panel).  The two histograms for the NIC2 field
represent the results for the entire frame (upper histogram) and the
corners of the frame only (lower histogram).  Here the ``corners'' are
defined as the region outside of an inscribed circle of radius
9\farcs6.  The bottom panel shows the ratio of the luminosity
functions for the entire NIC2 and NIC1 fields.  The solid horizontal
line is the overall count ratio for $\mbox{F110W}<18.4$ and the dotted
horizontal lines represent the 1-$\sigma$ errors on the ratio.
\label{fig:lf_comp}}
\bigskip
\end{figurehere}

The HB is clearly evident at $\mbox{F110W} = 15.9$ in all of the
histograms in Figure~\ref{fig:lf_comp}.  There is also a smaller
second peak at $\mbox{F110W} = 16.9$ evident in the NIC1 histogram.
This feature corresponds to the clump of stars seen below the HB in
Figure~\ref{fig:cmd_comp}, and appears to represent the RGB bump. To
estimate its statistical significance, we compare its height (11
stars) to the mean of the two adjacent bins (3 and 4 stars); the
result is a $2.1\sigma$ $t$-ratio.  Given that this is approximately
where the RGB bump should occur in a very high metallicity cluster, we
are fairly confident that we have detected it in the NIC1 field.  In
contrast, examination of the histogram for the entire NIC2 frame shows
no evidence of a RGB bump.  There is a hint of a RGB bump in the
histogram for stars in the frame corners, although this feature has
low statistical significance ($1.6\sigma$).  It may be that a small
RGB bump in the NIC2 field has been smeared out by photometric error.
The lack of a RGB bump in the histogram for the entire NIC2 field is
reflected by the dip in the NIC2/NIC1 count ratio at $\mbox{F110W}
= 16.9$.

The RGB bump detected in the NIC1 field provides an estimate of the
metallicity of Terzan~5.  The middle panel of Figure~\ref{fig:lf_comp}
indicates a value of $\dbump{\mbox{F110W}\,}=+1.0$.  We expect a
similar value to hold for $\dbump{V}$, since there is little color
difference between the RGB bump and the intersection of the HB with
the RGB\@.  As noted above, the largest $\dbump{V}$ observed by
\citet{fer99b} is +0.78 for NGC~6528.  For \dbump{V} = +1.0, their
$\dbump{V}-\zzin$ calibration gives $\zzin = +0.26$ for Terzan~5
\citep[where Z85 =][]{zin85}.  This is close agreement with the value
of +0.24 from \citet{zin85}, which is based on \emph{integrated}
infrared photometry.  It is also consistent with the conclusion by
\citet{ort96}, based on the RGB curvature, that the metallicity of
Terzan~5 is higher than that of NGC~6553
($\zzin=-0.29$)\footnote{Several recent determinations of the
metallicity of NGC~6553 have been carried out by \citet{bar99},
\citet{coh99}, and \citet{coe01}.  While there some differences among
these studies in detail, there is general agreement that the
metallicity of NGC~6553 is near solar.} and is probably solar.  Thus
there appears to be a strong indication that the metallicity of
Terzan~5 is at least solar and is perhaps even somewhat higher.  This
is particularly interesting, in light of our detection of an RR Lyrae
in the core \citep[see \S\ref{blue_stars} above]{edm01}, since this
would make Terzan~5 the highest metallicity cluster in which an RR
Lyrae has been observed.  This appears to provide yet another
indication of a high stellar interaction rate in the core of Terzan~5,
on the assumption that collisionally induced mass loss was responsible
for the production of the RR Lyrae.

\subsection{Reddening and Distance Modulus \label{distance}}

Figure~\ref{fig:nic1cmd} shows the CMD derived from the NIC1 frame
photometry, calibrated in the Johnson $JH$ magnitude system.  The
upper RGB is sparsely populated, given the small size of this field
and its $30\arcsec$ offset from the cluster center.  Nevertheless, the
six stars above the HB are sufficient to characterize the location
and slope of the upper RGB\@.  The dashed line represents a reddened
fit of the Baade's Window (BW) M-giant sequence from
\citet[Table~3B]{fro87}, based on the \citet{rie85} infrared reddening
law.  The inferred infrared reddening is $E(\jh) = 0.72$ and the total
infrared extinction is $A_H = 1.18$.  This corresponds to a visual
reddening of $\ebv=2.18$ and a $V$-band extinction of $A_V = 6.75$.
Our reddening estimate is somewhat smaller than the value of
$\ebv=2.49$ obtained by \citet{ort96} from the position of the RGB in
their ground-based $VI$ CMD\@.  Other previous reddening estimates
include $\ebv=1.65$ \citep[based on a spectroscopic measure of an
interstellar band]{arm88}, $\ebv=2.1$ \citep[based on integrated
infrared photometry]{mal82}, and $\ebv=2.42$ \citep[method
unspecified]{djo93}.  Inasmuch as the properties of Terzan~5 are best
determined in the infrared, it is interesting to note the close
agreement between our new reddening estimate and that of
\citet{mal82}.

\begin{figurehere}
\bigskip
\centering\includegraphics[width=3.2in]{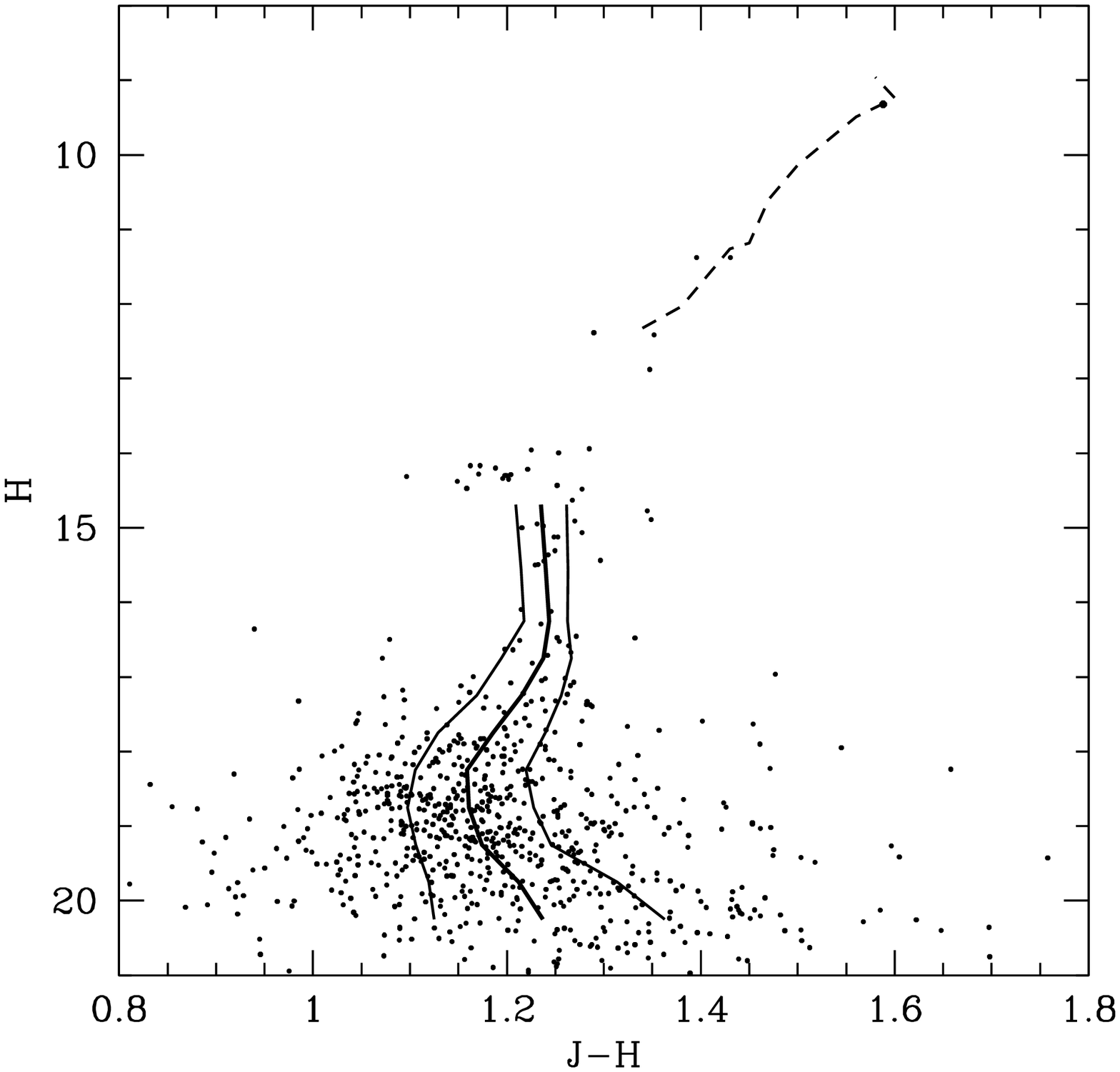}
\figcaption[f6.eps]{\protect\small CMD for the offset NIC1 field.  The
field center is located at 30\arcsec\ from the cluster center.  The
photometry has been transformed to the $JH$ system.  The dashed line
is a fit of the Baade's Window M-giant sequence, with $E(\jh)=0.72$.
The heavy solid curve is the CMD median ridge line and the lighter
solid curves represent the first and third quartiles of the color
distribution.
\label{fig:nic1cmd}}
\bigskip
\end{figurehere}

As we note in \S\ref{intro}, \citet{ort96} obtained a significantly
closer distance for Terzan~5 (5.6~kpc) than have other investigators.
If we simply replace their reddening value of $\ebv=2.49$ by our value
of $\ebv=2.18$, their distance estimate would be increased to 8.7~kpc.
Alternatively, we can estimate the distance by comparison with the
high-metallicity cluster NGC~6528, which has been studied in the
infrared by \citet{dav00}.  The reddening-corrected HB magnitude of
NGC~6528 is $H_{\rm HB}=13.1$ versus 13.0 for Terzan~5.  Adopting a
distance of 9.1~\kpc\ for NGC~6528 \citep{har96} also yields
$d=8.7~\kpc$ for Terzan~5.  Recent galactic center distance
determinations have produced values of $d_{\rm GC}=8.3\pm1.0~\kpc$
\citep{car95} and $d_{\rm GC}=7.9\pm0.3~\kpc$ \citep{mcn00}.  Since
the line of sight to Terzan~5 passes within 0.6~kpc of the galactic
center, the total distance of Terzan~5 from the galactic center
appears to be no more than about 1~kpc.  Uncertainties in both the
\nicmos\ calibration and the infrared reddening law contribute to the
error in the distance determination.  These uncertainties are not well
quantified.  We note that if the distance modulus is accurate to at
least 0.4~mag, then the uncertainty in the distance is less than about
2~kpc.

The apparently small distance of Terzan~5 from the galactic center is
interesting in light of the metallicity gradient in the bulge
predicted by dissipative collapse models for its formation
\citep*[e.g.][]{mol00}.  \citet*{fro99} have measured a metallicity
gradient of about $-0.45$~dex/kpc along the minor axis of the bulge.
\citet{fel00} have recently found a steeper gradient value of
$-1.3$~dex/kpc.  Based on near-infrared spectroscopy of bulge giants,
\citet{ram00} argue that there is little evidence for a gradient
\emph{inside} of 560~pc from the galactic center and find a mean
metallicity of $-0.21\pm0.30$ (s.d.) for the inner bulge.  Since the
metallicity that we find here for Terzan~5 is at least this large and
the bulge metallicity drops outside of the inner bulge, an inner bulge
origin for Terzan~5 appears to be a reasonable inference.

\subsection{The Main Sequence Turnoff and Cluster Age \label{age}}

As we noted in \S\ref{cmd}, it is evident from
Figure~\ref{fig:cmd_comp} (right panel) that the NIC1-field photometry
reaches below the main sequence turnoff (MSTO)\@.  The narrow 1.0 mag
color range used in the NIC1 $JH$ CMD shown in
Figure~\ref{fig:nic1cmd} emphasizes the spread of the points about the
mean color-magnitude relation.  To determine the location of the MSTO,
we computed the median color for 0.5 mag bins in $H$.  We also
computed the first and third quartiles of the color distributions in
each magnitude bin, in order to characterize the width of the
distributions.  These three color-distribution statistics are plotted as
continuous curves in Figure~\ref{fig:nic1cmd}.  The median color
reaches its bluest value near $H=18.5$, which is 4.3 mag below the
HB\@.  We identify this as the MSTO and note that its detection at
30\arcsec\ from the cluster center is an impressive demonstration of
the capability of NIC1 imaging.  The corresponding unreddened
magnitude and color of the turnoff are $H_{\rm TO}=17.3$ and
$(\jh)_{\rm\,TO}=0.4$.  The semi-interquartile range of the color
distribution at the turnoff is just 0.06 mag, indicating a fairly
narrow photometric spread for a typical star.

The magnitude difference between the HB and the MSTO, $\dmhbto{V}$, is
an important indicator of cluster age, as discussed by \citet{cha96}.
\citet{gir00} have recently computed evolutionary tracks for low- to
intermediate-mass stars with a wide metallicity range, and have
derived isochrones in the Johnson-Cousins system, including the $J$
and $H$ bands.  These can, in principle, be used to calibrate the
relation between $\dmhbto{H}$ and cluster age.  In applications of the
$\dmhbto{V}$ method of age determination, it is usual to adopt a
\emph{fixed} absolute magnitude for the HB\@.  This is typically the
RR Lyrae magnitude, which is assumed to be independent of age over the
relevant age range for globular clusters \citep[see
e.g.][]{cha96,ros99}.  For very high-metallicity clusters, where RR
Lyraes may be deficient or absent, an alternative choice for defining
a fixed HB magnitude, is the zero-age HB (ZAHB) magnitude for a star
of a chosen fudicial mass and appropriate metallicity \citep[see][as
discussed below]{ort01}.

As a cluster evolves, the magnitude of the ZAHB is expected to become
somewhat fainter, as indicated by the \citet{gir00} isochrones, which
are based on evolutionary tracks that include HB and post-HB
evolution.  The resulting evolution of the red giant clump (i.e.\ red
HB) absolute magnitude is shown by \citet[][see their Fig.~1]{gir01}.
Thus, an alternative procedure for calibrating the $\dmhbto{H}$-age
relation is to use the evolving HB magnitude.  In
Figure~\ref{fig:age}, we compare these two calibration procedures for
two values of the cluster metallicity.  In the upper panel, we use the
contemporaneous HB magnitude from the \citet{gir00} isochrones, while
in the lower panel we use the fixed, zero-age HB magnitude for a star
with an initial mass of 1\msun, following \citet{ort01}.  We define
the contemporaneous HB magnitude by the bluest point on the HB
isochrones, since examination of the \citet{gir00} isochrones
indicates that stars linger near this location.  It can be seen that
the calibration curves flatten considerably for $t \gtrsim 10~{\rm
Gyr}$ when the evolving HB magnitude is used.  In both panels, we show
the uncertainty range produced by an estimated 0.2~mag precision in
the determination of $\dmhbto{H}$.

\begin{figurehere}
\centering\includegraphics[width=2.8in]{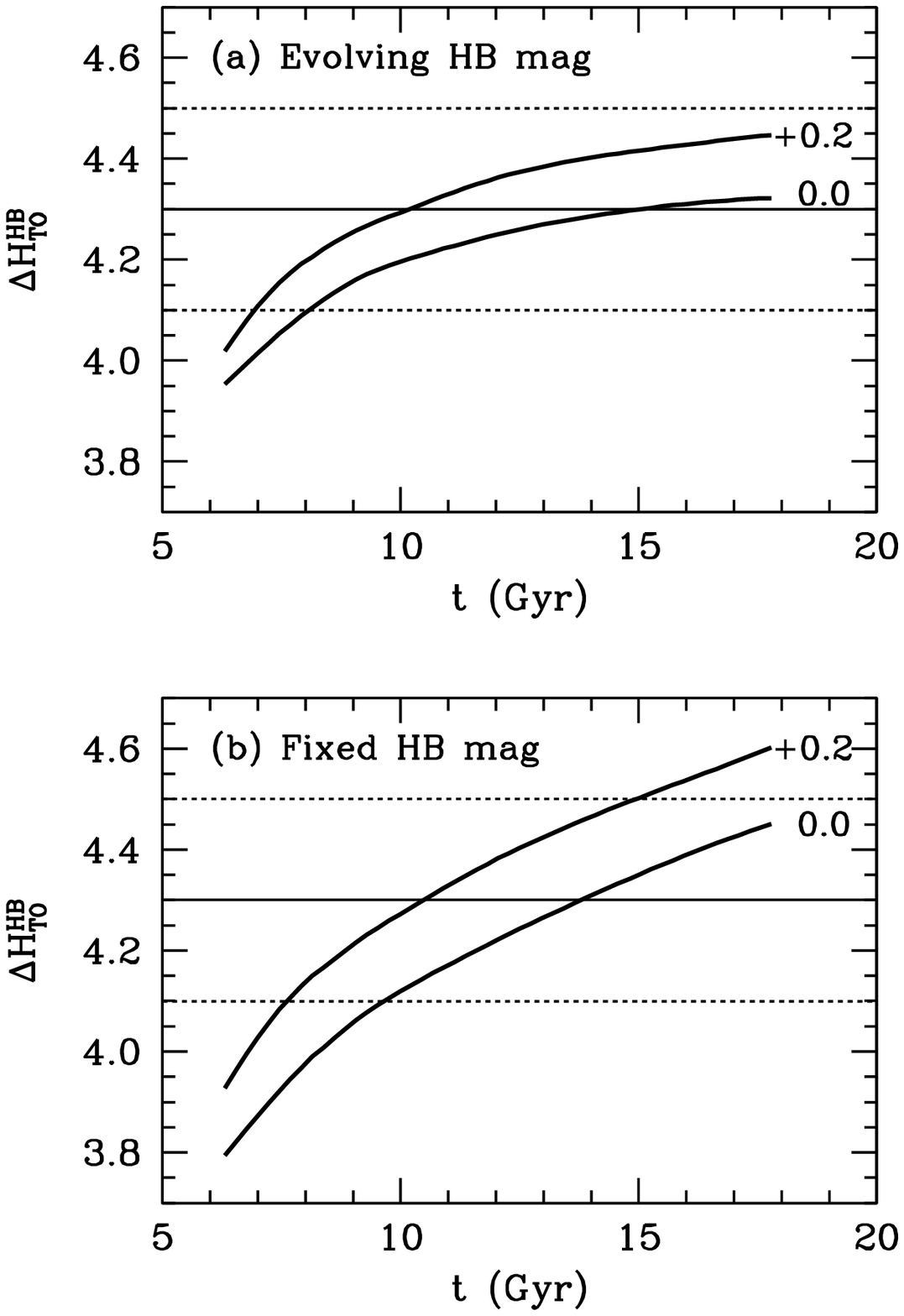}
\vspace*{-0.15in}
\figcaption[f7.eps]{\protect\small Magnitude difference between the
horizontal branch and turnoff, $\dmhbto{H}$, versus cluster age, for
(a) a time-evolving HB magnitude, and (b) a fixed HB magnitude.  In
each panel, the curves are labeled by the [Fe/H] value.  The solid
horizontal line in each panel is the observed value of $\dmhbto{H}$
for Terzan~5 and the dotted lines represent the estimated range of
uncertainty.
\label{fig:age}}
\end{figurehere}

The curves in the upper and lower panels of Figure~\ref{fig:age} give
similar results for the best-fit cluster age for a given value of the
metallicity.  However, it can be seen that the uncertainty ranges are
significantly different for the two calibration methods.  For an
evolving HB magnitude, the best-fit ages are 15~Gyr for solar
metallicity and 10~Gyr for supersolar metallicity, with lower limits
of 8~Gyr and 7~Gyr, respectively.  The flattening of the calibration
curves in this case precludes placing upper limits on the ages for the
estimated 0.2~mag precision of $\dmhbto{H}$.  The limits are tightened
significantly for a fixed HB magnitude, where the best-fit ages are
$14_{-4}^{+5}~{\rm Gyr}$ for solar metallicity and $11_{-3}^{+4}~{\rm
Gyr}$ for supersolar metallicity.

We note that \citet{ort01} have independently determined an age for
Terzan~5 by fitting isochrones to a CMD determined from their NIC1
observations of a field 30\arcsec\ from the cluster center.  They find
an identical offset of $4.3\pm0.2$~mag between the HB and the MSTO, at
F110W, to what we find at $H$, from which they infer an age of
$14\pm3$~Gyr, using newly computed isochrones in F110W--F160W HST
filter system.  They assume a fixed ZAHB magnitude, taken to be that
of a star of initial mass of 1~\msun.  They also adopt solar
metallicity.  Thus, they have treated the case represented by the
lower curve in the lower panel of Fig.~\ref{fig:age}, for which our
best-fit age is also 14~Gyr.  However, their computed values of
$\dmhbto{\mbox{F110W}}$ have a steeper dependence on age than do the
values for $\dmhbto{H}$ that we derived from the \citet{gir00}
isophotes, allowing \citet{ort01} to place somewhat tighter limits on
the determined age ($\pm3~{\rm Gyr}$ versus $\pm4.5~{\rm Gyr}$).  This
evidently results from a difference in the behavior of the isochrone
sets.  The good agreement between the best-fit values for the age of
Terzan~5 (assuming solar metallicity), obtained here and by
\citet{ort01}, is encouraging.  However, the assumption of a fixed HB
magnitude, when not using RR Lyraes to define the HB, should be
investigated further.

\section{Cluster Structure}

\citet*{tra95} have studied the spatial structure of Terzan~5 by
determining the surface-brightness profile from an $I$-band CCD image.
They determined a core radius value of $r_c=10\farcs7$ and a half-mass
radius of $r_h=55\arcsec$.  Since small numbers of individual bright
giants dominate the surface-brightness distribution, especially in the
$I$-band, star-count profiles generally provide a better means of
determining the cluster structure \citep[see e.g.][]{sos97}.

We have constructed a surface-density profile for Terzan~5, using star
counts from the central and offset NIC2 fields (see
Figure~\ref{fig:map}).  It is necessary to include the offset field,
since the cluster core fills the central field.  We used the F187W
star counts, since this is the only broad-band filter common between
these two fields.  Since our exposures for the offset field are not
dithered, it was not feasible to perform PSF-subtraction photometry
for these frames.  Instead, we used simple aperture photometry and
computed a magnitude offset between the central and offset fields from
the stars in the overlap region.  Since the depth of the star counts
varies with distance from the cluster center, we adopted a uniform
limiting magnitude for stars to be included in the profile
determination that corresponds to 2 mag below the mean HB, i.e.\ F187W
= 15.9.  Figure~\ref{fig:lf_comp} indicates that the central field is
fairly complete to this limit.

\begin{tablehere}
\begin{deluxetable}{lllc}
\tablecaption{Comparison of Cluster Center Determinations. \label{tbl:centers}}
\tablewidth{0pt}
\tablehead{
\colhead{Study} & 
\colhead{$\alpha$ (2000)} &
\colhead{$\delta$ (2000)} &
\colhead{$\Delta~('')$~\tablenotemark{a}}
}
\startdata
\citet{ter71}   & 17 48 04.6  & $-$24 46 50.9 & 6.7 \\
\citet{dm93}\tablenotemark{b} 
                & 17 48 04.9   & $-$24 46 45   & 1.1 \\
\citet{pic95}   & 17 48 04.6   & $-$24 46 48.5 & 4.7 \\
\citet{fru00}   & 17 48 04.9   & $-$24 46 44.7 & 1.3 \\
Present study   & 17 48 04.8   & $-$24 46 45.0 & --- \\
\enddata

\tablenotetext{a}{Total offset relative to present study.}
\tablenotetext{b}{A misprint in the $\delta$ minutes value has been
corrected per private communication with S.~Djorgovski.}

\tablecomments{All positions are given with the same precision as in
the original source.}

\end{deluxetable}
\end{tablehere}

We used iterative centroiding, in a circular aperture with an initial
radius equal to the half width of the NIC2 field (9\farcs6), to
determine the location of the cluster center.  As the centroid shifted
from the starting position at the frame center, we reduced the size of
the aperture to keep it entirely within the central NIC2 field at each
iteration.  The iteratively determined centroid was offset by
$0\farcs7$ from the frame center.  Trials with a range of values of
the limiting magnitude and aperture size indicate that the uncertainty
in the location of the center is of order 1\arcsec.  Our determined
center position corresponds to the equatorial coordinates (J2000)
$\alpha = 17^{\mathrm h}~48^{\mathrm m}~04\fs8$ and $\delta =
-24\arcdeg~46\arcmin~45$.  Table~\ref{tbl:centers} compares this
position with previous determinations.  There is particularly good
agreement between the present center determination and those of
\citet[][after correction; see Table~\ref{tbl:centers} notes]{dm93}
and \citet{fru00}.  The latter is the centroid of the unresolved radio
flux from the core, which they suggest arises from an unresolved MSP
population.  The close agreement between this radio center and our
present optical center is striking.

We binned the star counts into a central disk of radius 0\farcs75 and
nine concentric, logarithmically spaced circular annuli that extend to
29\arcsec\ from the cluster center; the annuli are incomplete beyond
9\arcsec.  The surface density in each annulus is computed by dividing
the number of stars in that annulus by the area of the annulus that
lies within the combined NIC2 central and offset fields.
Figure~\ref{fig:profile} shows the resulting surface-density profile.
We have included the \citet{tra95} surface-brightness profile for
comparison\footnote{The \citet{tra95} data set includes three spectral
bands, $B$, $R$, and $I$.  The $I$ and $R$ data cover the entire
profile, while the $B$ data only cover the outer halo.  Since the $I$
and $R$ data each tightly define a profile that is slightly different
from that defined by the other band, we chose to include only their $B$
and $I$ data in this comparison.}, with a shift applied to it so that
the two profiles agree near $r=10\arcsec$.  It can be seen that both
profiles shown in Figure~\ref{fig:profile} have relatively flat cores,
with a somewhat smaller core radius for our surface-density profile.

In order to determine the core radius in a manner independent from the
adopted binning, we employed the maximum-likelihood method outlined by
\citet{sos97}.  In this approach, a surface density profile function
is fit to the star position data directly, without resort to radial
binning.  The likelihood for the observation of a particular set of
stellar radial position values ($r_i$) is given by:
\beq
L(\mathbf{a}) = \prod_{i=1}^N f(r_i;\mathbf{a})   \label{eqn:Likelihood}
\eeq
where $f$ represents a parameterized form for the surface density
profile (in units of stars per unit area) and $\mathbf{a}$ represents
a set of parameters on which $f$ depends.  It is convenient to work
with $\log L$, which converts the product in
equation~(\ref{eqn:Likelihood}) to a sum.  Rather than compute $\log
L$ over a grid in parameter space, to locate the point of maximum
likelihood, we found it more convenient to use an optimization
routine.  We used an implementation of Powell's method from
\emph{Numerical Recipes} \citep[{\S}10.5]{pre92}, which does not
require the computation of derivatives of the likelihood function with
respect to the parameters.  For estimating the confidence ranges for
the parameter estimates, we used the ``bootstrap'' method, i.e.\
random resampling of the star list \emph{with replacement}
\citep[{\S}15.6]{efr82,pre92}.  We implemented this by drawing 1000
random resamples and determining the maximum-likelihood values of the
parameters for each resample.  We verified that the distributions of
the parameter values determined from the set of resamples are closely
approximated by Gaussians, with means equal to the best-fit parameter
values for the original sample.  The standard deviations of these
distributions provide robust estimates for the uncertainty of the
best-fit parameter values.

\begin{figurehere}
\bigskip
\centering\includegraphics[width=3.2in]{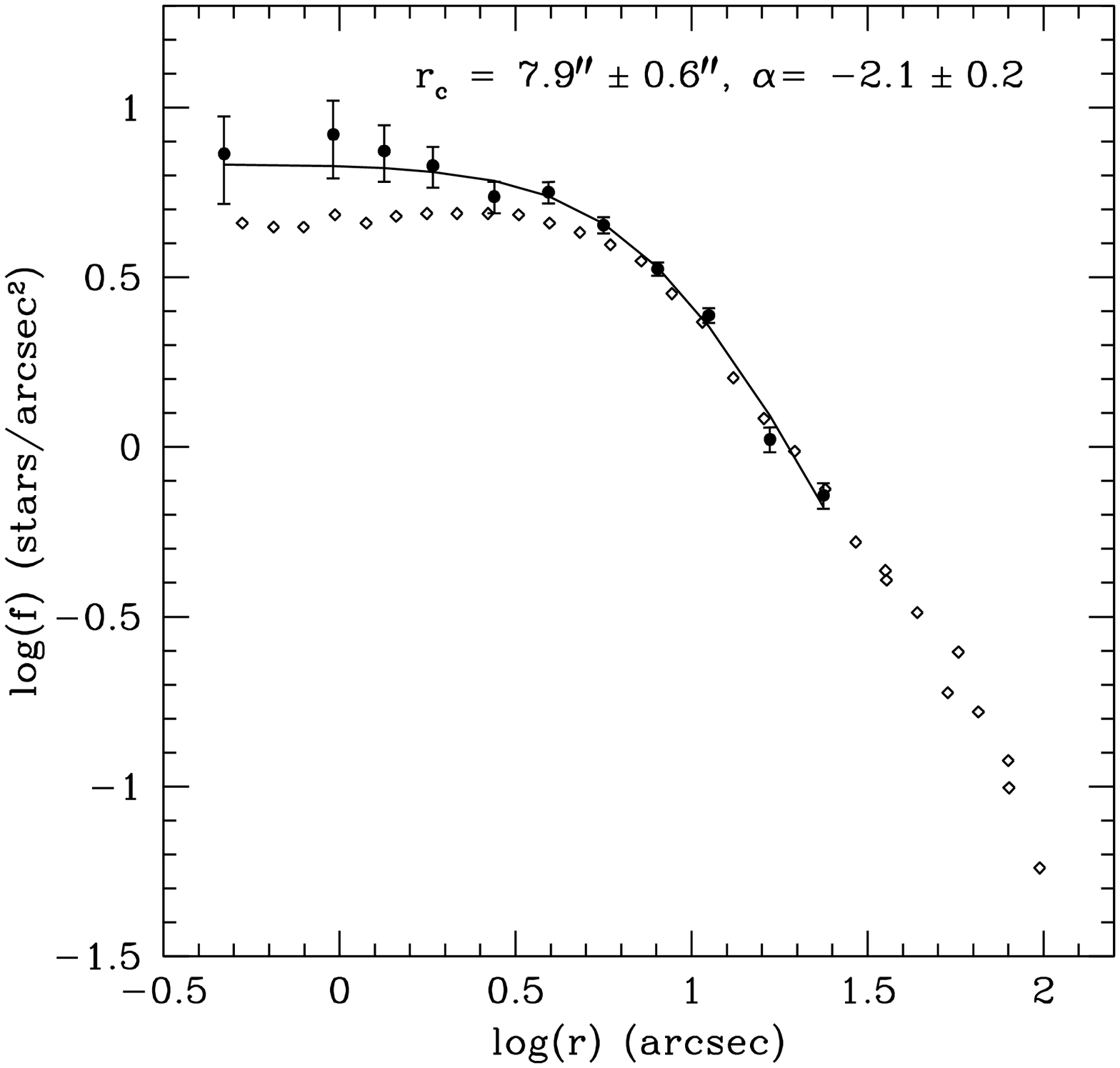}
\figcaption[f8.eps]{\protect\small Surface-density
profile from F187W star counts (solid dots and error bars) compared
with surface brightness profile from \citet[][open diamonds]{tra95}.
The solid line is a maximum likelihood fit of eqn.~(\ref{eqn:SDform})
to the \emph{unbinned} F187W star counts.
\label{fig:profile}}
\bigskip
\end{figurehere}

Following \citet*{lug95}, we used the following modified-power-law
form for the azimuthally symmetric surface density distribution,
\beq
f(r) = f_0 \left[ 1 + \left(r \over r_0 \right)^2 \right]^{\alpha/2}.
     \label{eqn:SDform}
\eeq
Here $r_0$ is a radial scale related to the core radius and $\alpha$
is the power-law index of the outer part of the profile.  The
normalization parameter $f_0$ is determined by the condition that the
integral of $f(r)$ over the area in which the stars are located
remains constant as the values of the other two parameters are varied.
For the usual definition of the core radius as the point at which the
surface density decreases to half of its central value, $r_c =
(2^{-2/\alpha}-1)^{1/2}\,r_0$.  A power-law slope of $\alpha=-2$
corresponds to an ``analytic King model'' \citep{kin62}.
\citet{lug95} have shown that equation~(\ref{eqn:SDform}) with
$\alpha=-1.8$ provides an excellent fit to an actual \citet{kin66}
model with a central concentration of $c=2.0$, for all but the outer
halo.  As a consistency check, we performed a least-squares fit of
equation~(\ref{eqn:SDform}) to the \citet{tra95} profile.  We find
$r_c=10\farcs2$ and $\alpha=-1.9$.  This core radius value agrees with
their value of 10\farcs7 to within 5\%; the power-law slope is
consistent with a King model.

Our maximum likelihood fit of equation~(\ref{eqn:SDform}) to the NIC2
star counts produces $r_c=7\farcs9\pm0\farcs6$ and
$\alpha=-2.1\pm0.2$.  It can be seen in Figure~\ref{fig:profile} that
this form provides an excellent fit.  The core radius is about 25\%
smaller than that found by \citet{tra95}, while the power-law slope is
similar to that produced by our fit to their profile.  Interestingly,
in a preliminary report of their profile fits, \citet*{tra93} found a
core radius of $7\farcs9$, in agreement with our value.  The slope value
of $\alpha \approx -2$ again indicates consistency with a King model.

\begin{figurehere}
\bigskip
\centering\includegraphics[width=3.2in]{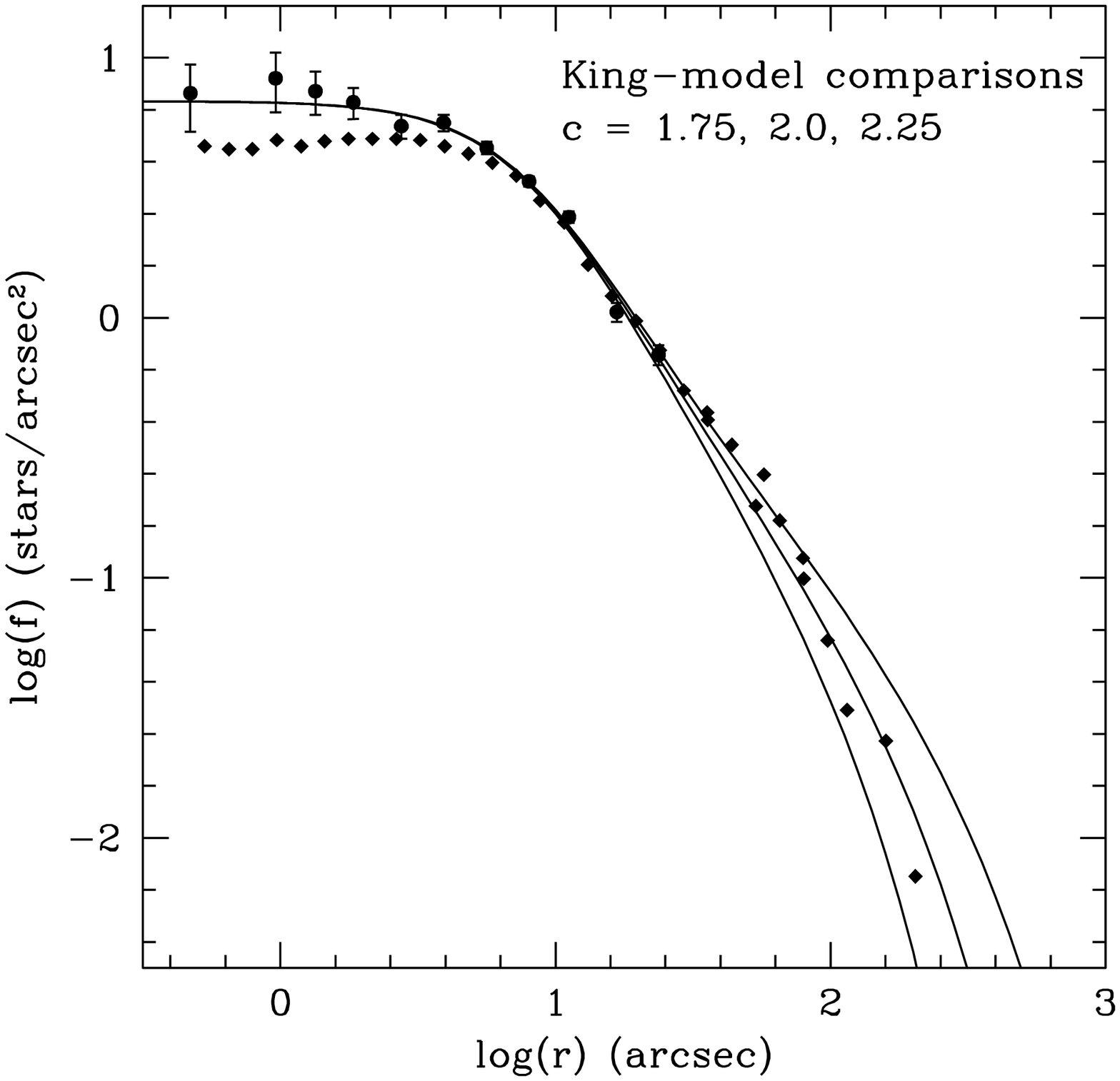}
\figcaption[f9.eps]{\protect\small Combined surface-density profile
(as described in Fig.~\ref{fig:profile}) with King model overlays.
The curves correspond to the indicated $c$ values from left to right.
The core radius and central density are taken from the maximum
likelihood fit shown in Fig.~\ref{fig:profile}.  Note that the best
match is obtained for concentration parameter in the range $2.0
\lesssim c \lesssim 2.25$.
\label{fig:King-models}}
\bigskip
\end{figurehere}

To further explore the limits on the central concentration of Terzan~5
that can be obtained from model fitting to the combined profile shown
in Figure~\ref{fig:profile}, we have overplotted a set of King models
with a range of $c$ from 1.75 to 2.25, as shown in
Figure~\ref{fig:King-models}.  Rather than formally fitting the
profiles, we simply adopted the core radius and central surface
density values from our maximum-likelihood fits of the modified
power-law profile.  It can be seen in Figure~\ref{fig:King-models}
that the $c=2.0$ King model provides the best overall fit to the other
part of the profile as determined by \citet{tra95}.  Given the
challenge of determining the surface brightness profile at large
distance from the cluster center, with the high background of the
galactic bulge and variable extinction, we do not view this as a
rigorous determination of the central concentration of Terzan~5.
Rather, we see it as a demonstration that a relatively high central
concentration of $c\approx2$ is certainly plausible.

For our adopted normalization of the profiles near $r=10\arcsec$, the
central surface density is 60\% higher for the NIC2 profile than for
that of \citet{tra95}.  The combination of the higher surface density
and the smaller core radius implies that the central density is about
twice as large as the value of $\rho_0 = 10^{5.76}~\msun\,\pc^{-3}$
determined by \citet{djo93}, i.e.\ $\rho_0 \approx
10^{6.0}~\msun\,\pc^{-3}$.  This puts Terzan~5 at the top of the
central density distribution for all clusters, although the tabulated
values for collapsed-core clusters are more properly viewed as lower
limits.  Following \citet{djo93}, we may estimate the central
relaxation time of Terzan~5, by applying the scaling relation, $t_{\rm
rc} \propto {\rho_0}^{0.5} r_c^3$, to his estimate.  The result is
$t_{\rm rc} = 4\times10^7~{\rm yr}$, which is at about the 30th
percentile of the distribution for 124 clusters from \citet{djo93}.
Of the clusters with shorter central relaxation times than Terzan~5,
only about one third have apparently ``normal'' cores, i.e.\ have a
central structure that is well fit by a King model, like that of
Terzan~5.  Thus, Terzan~5 has both an extremely high central density
and an unusually short relaxation time for a cluster with a
noncollapsed core.

\section{Discussion}

Using NIC2 imaging of the central $19\arcsec\times19\arcsec$ region of
Terzan~5 and NIC1 imaging of a field 30\arcsec\ from the cluster
center, we have constructed color-magnitude diagrams and produced new
measurements of the cluster reddening, distance, and metallicity.  We
have also determined the surface-density profile and performed a
maximum-likelihood fit to measure the core radius and halo slope.  Our
results underscore the extraordinary nature of this cluster.

The result that the reddening is somewhat smaller than found by
\citet{ort96} places the cluster at a greater distance than the
strikingly small value of 5.6~kpc that they found.  This new
reddening, together with our measurement of the $J$-band magnitude of
the horizontal branch, place Terzan~5 at the 60\% larger distance of
8.7~kpc and thus within less than 1~kpc of the galactic center.  Our
distance measurement is consistent with the range of 7 to 9~kpc
produced by most studies prior to \citet{ort96}.  

The probable detection of the red giant branch bump, about 1 mag
below the horizontal branch, indicates that the metallicity of
Terzan~5 is at least solar and possibly somewhat higher ($\zzin =
+0.26$).  This places Terzan~5 at the very top of the metallicity
distribution of all globular clusters.  This suggests that Terzan~5
was formed in the vicinity of its present location near the galactic
center.

The detection of the main sequence turnoff has allowed us to estimate
the cluster age using the magnitude difference between the turnoff and
the horizontal branch.  Our result for the best-fit age---14~Gyr for
an assumed solar metallicity and a fixed HB magnitude---is the same as
that recently obtained independently by \citet{ort01} from a separate
\nicmos\ data set.  However, our derived uncertainty in the age is
somewhat larger than they obtained, owing to differences in the
isochrone sets used.  Even with their tighter age precision of
$\pm3~{\rm Gyr}$, it is only marginally possible to distinguish
between ``young'' ($\sim10~{\rm Gyr}$) and ``old'' ($\sim15~{\rm
Gyr}$) ages for Terzan~5.  We also note the role of metallicity in the
age determination; Fig.~\ref{fig:age} indicates that the inferred age
may be reduced by 3-5 Gyr if supersolar metallicity were adopted.
Given the extraordinarily high metallicity of Terzan~5, a tighter
limit on the cluster age would have important implications for the
chemical history of the galactic center region, particularly if the
age turned out to be fairly old.

Improvements in the age estimate could be achieved by a combination
of: (1) NIC1 mosaic imaging of the central region to improve the
photometric precision and depth, (2) additional \nicmos\ imaging at
about 0.5--1\arcmin\ from the cluster center, to improve the
statistics of the deep CMD, (3) \nicmos\ observations of more
near-infrared standard stars, to improve the determination of the
magnitude system transformations, (4) additional study of infrared
isochrones to better fit the observations of high-metallicity systems,
and (5) a tighter limit on the metallicity of Terzan~5.

Our maximum-likelihood fit of a modified-power-law form to the
surface-density profile within 0\farcm5 of the cluster center has
produced a new core radius measurement of $r_c=7\farcs9$, which is
somewhat smaller than the previous determination by \citet{tra95}.
Nevertheless, our results underscore the fact that Terzan~5 has a
resolved core and is well fit by a King-model profile.  Given the
relatively short central relaxation time of $t_{\rm rc} =
4\times10^7~{\rm yr}$, it is interesting to ask whether Terzan~5
should have already undergone core collapse.  For a single-mass
cluster, the core collapse time is formally about 300 times longer
than the central relaxation time \citep{coh80}.  This would suggest
that Terzan~5 may be just now at the brink of core collapse \citep[see
also][]{coh84}.  Furthermore, \citet{mey97} have argued that any
cluster with central concentration $c \gtrsim 2$ should be considered
to be either on the brink of core collapse or just past it.  

While the concentration parameter of Terzan~5 has not been directly
determined, given the difficulty of determining the tidal radius of
this highly obscured cluster, our comparisons with King models suggest
that $c \approx 2$.  Given the estimated half-mass radius of Terzan~5
from \citet{tra95}, the ratio of core to half-mass radius is $r_c/r_h
= 0.15$.  \citet{gra92} have shown that this ratio has a maximum value
of about 0.01 -- 0.04 in post-collapse dynamical models for clusters.
This suggests that the core of Terzan~5 may still be supported in a
pre-collapse state by energy release (``heating'') from primordial
binaries that interact dynamically with other stars.  Alternatively,
Terzan~5 may be in a post-collapse state with a ``super-expanded''
core, produced by the action of a residual population of primordial
binaries.  Such binaries were not considered by \citet{gra92}.

If Terzan~5 is indeed near the point of core collapse, we might expect
to see enhanced binary activity in its core.  \citet{fer99a} have
interpreted the huge blue straggler population that they detected in
the high-density, noncollapsed core of M80 as evidence of vigorous
binary heating, which is temporarily staving off imminent collapse.
Specifically, they propose that the blue stragglers are the result of
the hardening of both primordial binaries and binaries produced by
tidal capture and three-body interactions.  The vast majority of blue
stragglers present in Terzan~5 should lie in the central NIC2 field of
our data set, given the extreme central concentration of blue
straggler populations in dynamically evolved clusters.  The likely
eclipsing blue straggler identified by \citet{edm01} lies at about
8\arcsec\ from the cluster center.  Highly dithered NIC1 photometry of
the central region may aid in detecting additional blue stragglers in
the core, although distinguishing between blue stragglers and
photometric artifacts in this crowded central region could prove
challenging.  Ground-based adaptive optics observations with the
\emph{Hokupa'a} system on \emph{Gemini} may well prove superior to
\nicmos\ for probing the core of Terzan~5, given the large telescope
aperture, the 0\farcs02 pixel size of the imager, the 20\arcsec\ field
of view, and the 0\farcs07 or better expected FWHM of the images under
good seeing conditions.

Multi-wavelength studies of Terzan~5, including additional searches
for millisecond pulsars and X-ray sources, will certainly be useful
for assessing the level of stellar collisional activity in the core.
Our group has recently obtained a deep \emph{Chandra} observation of
Terzan~5.  However, the LMXB was in a high state during this
observation, which may preclude the detection of a large population of
faint X-ray sources in the central region.  In any case, the X-ray
spectral observations should provide a measurement of the hydrogen
column density towards Terzan~5, and thus an independent estimate of
the obscuration.  This will provide a check on our new distance
estimate.  Our \nicmos\ study underscores the extraordinary physical
conditions present in the core of Terzan~5.  Thus, it remains of great
interest to test the prediction that this region is a factory for the
production of substantial numbers of interacting binaries.

\acknowledgments

We acknowledge helpful conversations with Charles Bailyn, Adrienne
Cool, Con Deliyannis, and Ata Sarajedini.  We also appreciate the
insightful review by the anonymous referee, which improved the
manuscript.  This research was supported in part by NASA grants
GO-07889.01-96A to Harvard University and GO-07889.03-96A to Indiana
University.


\end{document}